\documentclass[preprint,amssymb,preprintnumbers,amsmath,prl,aps,showpacs]{revtex4}
\usepackage{epsfig,dsfont,amssymb,amsmath,amsthm,amsfonts,amsbsy,mathrsfs}
\usepackage{graphicx}
\usepackage{amsmath}
\usepackage{amssymb}%

\begin{document}

\newcommand{\dx}{\mbox{${\bf d}_x$}}
\newcommand{\dy}{\mbox{${\bf d}_y$}}
\newcommand{\dz}{\mbox{${\bf d}_z$}}
\newcommand{\rx}{\mbox{${\bf r}_1$}}
\newcommand{\ry}{\mbox{${\bf r}_2$}}
\newcommand{\bN}{\mbox{${\bf N}$}}
\newcommand{\bP}{\mbox{${\bf P}$}}
\newcommand{\Ex}{\mbox{${\bf E}_x$}}
\newcommand{\Ey}{\mbox{${\bf E}_y$}}
\newcommand{\Ez}{\mbox{${\bf E}_z$}}
\newcommand{\dux}{\mbox{${\bf u}_1$}}
\newcommand{\duy}{\mbox{${\bf u}_2$}}
\newcommand{\drx}{\mbox{${\bf r}_1$}}
\newcommand{\dry}{\mbox{${\bf r}_2$}}
\newcommand{\strain}{\mbox{\boldmath $\gamma$}}
\newcommand{\andsp}{\mbox{$\quad\textrm{and}\quad$}}
\newcommand{\pfrac}[2]{\frac{\partial #1}{\partial #2}}
\newcommand{\Dfrac}[2]{\frac{\mathrm{d} #1}{\mathrm{d} #2}}

\newcommand{\ud}{\mathrm{d}}

\title{Nonlinear geometric effects in mechanical bistable morphing structures}

\author{Zi Chen$^{1,2,4}$, Qiaohang Guo$^{5,6}$, Carmel Majidi$^{8}$, Wenzhe Chen$^{5,7}$, David J. Srolovitz$^{9}$, Mikko P. Haataja$^{1,2,3,10}$}
\affiliation
{$^1$Dept. of Mechanical and Aerospace Engineering, \\
$^2$Princeton Institute for the Science and Technology of Materials (PRISM),\\
$^3$Program in Applied and Computational Mathematics, Princeton University, Princeton, NJ 08544, USA\\
$^4$Dept. of Biomedical Engineering, Washington University, St. Louis, MO 63130\\
$^5$College of Materials Science and Engineering, Fuzhou University, Fuzhou 350108, China\\
$^6$Dept. of Mathematics and Physics, Fujian University of Technology, \\
$^7$Dept. of Materials Science and Engineering, Fujian University of Technology, Fuzhou 350108, China\\
$^8$Dept. of Mechanical Engineering, Carnegie Mellon University, Pittsburgh, PA 15213, USA\\
$^9$Institute for High Performance Computing, 1 Fusionopolis Way, Singapore 138632\\
$^{10}$School of Mathematics, Institute for Advanced Study, Princeton, NJ 08540, USA}


\begin{abstract} 
Bistable structures associated with non-linear deformation behavior, exemplified by the Venus flytrap and slap bracelet, can switch between different functional shapes upon actuation. Despite numerous efforts in modeling such large deformation behavior of shells, the roles of mechanical and nonlinear geometric effects on bistability remain elusive. We demonstrate, through both theoretical analysis and table-top experiments, that two dimensionless parameters control bistability. Our work classifies the conditions for bistability, and extends the large deformation theory of plates and shells.

\end{abstract}%



\maketitle


Mechanical structures exhibiting multiple stable morphologies arise in a variety of physical systems, both natural \cite{Forterre_Nature2005} and synthetic \cite{Hyer_1981,Daynes_2010,Kebadze_2004,Vidoli_ProcRSocA2008}.
For example, the leaves of the Venus flytrap can be triggered to snap-through from an open to a closed state in a fraction of a second to capture insects 
 \cite{Forterre_Nature2005}. Indeed, multi-stable components have promising applications in micropumps, valves, shape-changing mirrors in adaptive optical systems, mechanical memory cells \cite{Vidoli_ProcRSocA2008}, artificial muscles, bio-inspired robots \cite{Shahinpoor_2011}, deployable aerospace components \cite{Daynes_2010}, and energy harvesting devices \cite{Arrieta_APL2010}.

Multi-stability of plates and shells associated with non-linear deformation has inspired many studies over the years
\cite{Hyer_1981,Daynes_2010,Kebadze_2004,Vidoli_ProcRSocA2008,Finot_JAP1997,Galletly_IntJSolStruct2004II,Guest_ProcRSocA2006,Seffen_ProcRSocA2007,Gigliotti_2004,Fernandes_2010,Armon_Science2011,Forterre_Science2011}, while related bifurcation phenomena in slender structures \cite{DomokosHealey_2004}, such as DNA \cite{Biton_2007, Kocsis_2012} and plant tendrils \cite{Goriely_1998}, have also been long investigated \cite{Huang_2012}. Existing models of shell structures utilize geometric assumptions or scaling arguments that greatly simplify analysis, leading to solutions that have significantly improved our understanding of bistability.  Nonetheless, these geometric assumptions are typically only valid for small angle deformations and often do not strictly satisfy the mathematical conditions for geometric compatibility \cite{Docarmo_DiffGeo}, especially in  large deformation cases. Scaling theory has been employed by Forterre et al.~\cite{Forterre_Nature2005}, who proposed a minimal model to explain some of the bistable features of Venus flytraps, with a dimensionless deformation energy postulated {\it in lieu} of a ``first-principles'' derivation, and Armon et al.~\cite{Armon_Science2011}, who investigated the competition between bending and stretching deformation within incompatible thin sheets.  In the latter work, a dimensionless width was proposed to be the key bifurcation parameter, with the threshold value obtained through experiments. While this study is elegant, it did not yield quantitative predictions for the bifurcation threshold; furthermore, the presence of an {\it a priori} unknown quantity (curvature) in the bifurcation parameter further restricts its utility.  Therefore, for large deformations of shells, a more general elasticity theory framework, that explicitly includes the non-uniform bending curvature and mid-plane stretching, is required.

In this work, we propose an analytically tractable, reduced-parameter model that incorporates geometric nonlinearity and accounts for the competition between bending and in-plane stretching involved when a planar sheet is mapped into a non-developable shape via misfit strains.
We employ this model to perform a comprehensive examination of bistability, leading to unique predictions for bifurcation thresholds and bistable morphologies that are quantitatively validated with a series of table-top experiments.  More generally, our approach provides a quantitative means to investigate the role of geometric nonlinearities in a broader setting, including morphogenesis \cite{Liang_PNAS2009,Klein_Science2007,Dervaux_PRL2008, Wyczalkowski2012}, film-on-substrate electronics \cite{Suo_APL1999,Chen_JAM2004,Rogers_PNAS2009}, and spontaneous bending, twisting and buckling of thin films and helical ribbons \cite{Chun_nanolett2010,Zhang_nanolett2006,Chen_APL2011,Sawa_PNAS2011,Shenoy_ACSNano2010}.

\begin{figure}[t]
\centerline{\includegraphics[height=3.25 in]{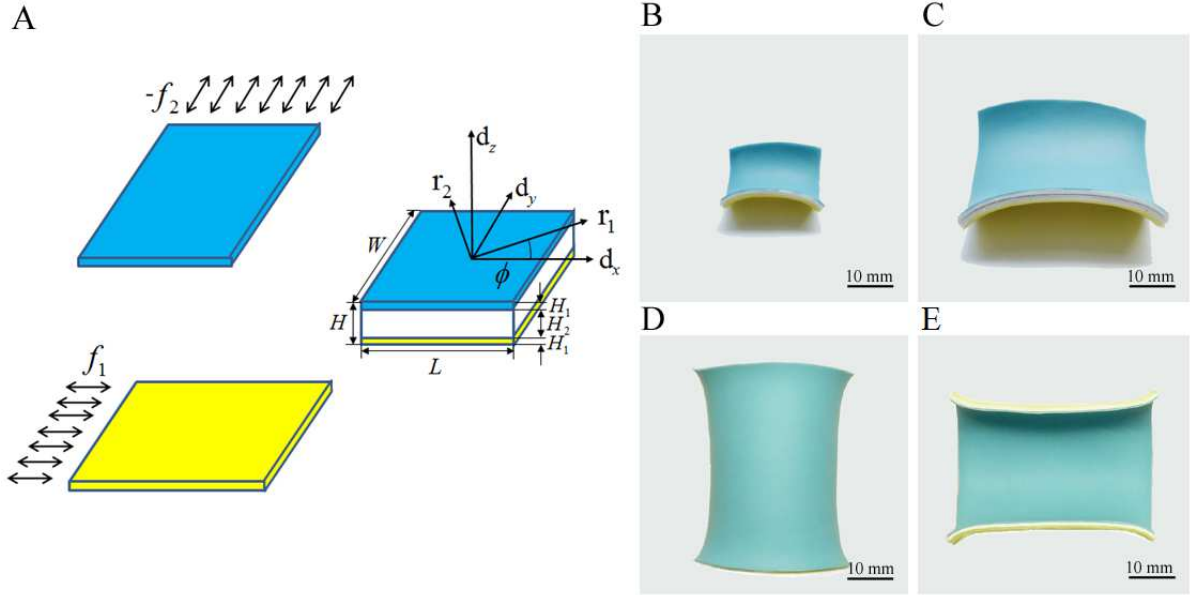}}
\caption{(A) Schematics: two thin latex rubber sheets (blue and yellow) were pre-stretched along perpendicular directions and bonded to a much thicker elastic strip. When released, the bonded multilayer sheet deforms into one of the following shapes.  Here, $H$ denotes the overall thickness, while angle $\phi$ measures the misorientation of the principal axes of curvature (${\bf{r}}_1$ and ${\bf{r}}_2$) from the principal geometric ones (${\bf{d}}_x$ and ${\bf{d}}_y$, respectively). (B) A saddle shape for a small, thin square sheet ($H = 1.5$mm, $L = W = 24.0$mm). (C) A saddle shape for a thick square sheet ($H = 2.5$mm, $L = W = 48.0$mm).  (D) A stable, nearly cylindrical shape (curving downwards) for a thin, wide  strip ($H = 1.5$mm, $L = W = 48.0$mm). (E) The other stable, nearly cylindrical configuration (bending upwards) for the same sheet as in (D). The nearly cylindrical shape smoothly transitions into a doubly curved shape near the edges.}
\label{fig:Exp1}
\end{figure}

We employ a simple experimental demonstration of the morphological transition from mono- to bi-stable states when the dimensions change (either the width increases or thickness decreases), with all other parameters fixed. To this end, two pieces of thin latex rubber sheets (thickness $H_1$, length $L$, and width $W \leq L$) are pre-stretched by an equal amount and bonded to an elastic strip of thicker, pressure-sensitive adhesive (thickness $H_2$) \cite{Footnote1} as shown schematically in Fig.~\ref{fig:Exp1} A, such that the total thickness of the bonded strip is $H = 2H_1 + H_2$. The pre-strains along the two lateral directions were chosen to be equal to $0.28$ in all of the experimental results presented.  When released, the initially flat, bonded laminate deforms into either a saddle shape (Fig.~\ref{fig:Exp1} B and C) or one of two nearly cylindrical configurations (Fig.~\ref{fig:Exp1} D and E), driven by misfit strains. More specifically, when the strip is sufficiently narrow or thick, the equilibrium saddle shape is unique; while if the strip is wide and/or thin, it will bifurcate into one of the two nearly cylindrical shapes.

As the starting point of our theoretical approach, we assume that an originally flat elastic sheet of $H \ll W \leq L$, with a rectangular cross-section and principal geometric axes along its length ($\dx$), width ($\dy$), and thickness ($\dz$), deforms into part of a torus shape with variational parameters $\kappa_1$ and $\kappa_2$ ($\kappa_i = 1/R_i$, $i$ = 1,2) along principal directions ${\bf r}_1$ and ${\bf r}_2$ (which coincide with the geometric axes) as shown in Fig.~2. A torus can describe a broad range of possible morphologies, such as a saddle, cylinder, and sphere; selected by tuning two geometric shape parameters. 
Instead of assuming small deformation as in classical lamination theory, we allow for large 
deformations with the associated geometric nonlinearity in the small elastic strain limit (in the thick sheet); that is, we only require that \textbf{$\kappa_i W \ll 1$}. Finally, mapping the sheet onto the surface of a torus (or part thereof) facilitates the explicit construction of the strain tensor components \cite{note}: $\gamma_{xx}(t,z) = [\cos(\kappa_2 t)z]/ \left\{1/\kappa_1 + [\cos(\kappa_2 t) - 1]/\kappa_2 \right\}  + [\cos(\kappa_2 t) - 1]\kappa_1/\kappa_2$, $\gamma_{yy}(t,z) = \kappa_2 z $, and $\gamma_{zz}(t,z) = -\nu (\gamma_{xx} + \gamma_{yy})/(1-\nu)$, where $t$ and $z$ denote the arclength along ${\bf r}_2$ and  the distance from the mid-plane of the shell.


\begin{figure}[t]
\centerline{\includegraphics[height=3.75 in]{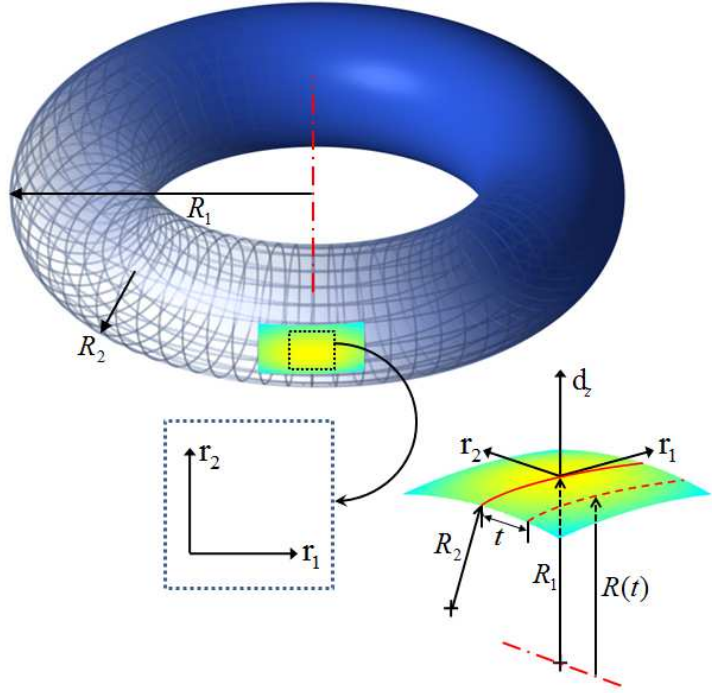}}
\caption{Deformation of an originally flat elastic strip into a section of a torus with variational parameters $\kappa_i$ ($\kappa_i = 1/R_i$, i = 1,2). Here, ${\bf r}_1$ and ${\bf r}_2$ denote the principal bending axes and $t$ denotes the arclength measured from the circle of radius $R_1$. The red dotted line at $t$ 
is shorter than the red solid line, indicating an extra 
strain due to geometric constraints.}
\label{Torus}
\end{figure}

The strip is subjected to effective surface stresses (due to the thin, misfitting top and bottom layers) of the form ${\bf f}^{+}= -f_2 \dy \otimes \dy$ and ${\bf f}^{-}=  f_1 \dx \otimes \dx$ along the top and bottom surfaces (``$\otimes$'' denotes the tensor direct product), respectively. 
This is equivalent to the case in which a strip is subjected to only one surface stress ${\bf f}^{-} =  f_1 \dx \otimes \dx + f_2 \dy \otimes \dy$ on the bottom surface, where only the bending mode of deformation is of interest \cite{Chen_APL2011}. For conciseness, we assume that the principal axes of bending, ${\bf r}_1$ and ${\bf r}_2$, coincide with the principal axes of the surface stresses, i.e.,  the geometric axes, $\dx$ and $\dy$ (justified below).

Continuum elasticity theory gives the potential energy density per unit length of the strip as $\Pi = \int_{-W/2}^{W/2} \Big[{\bf f}^-:\strain |_{z=-H/2} + \int_{-H/2}^{H/2}\frac{1}{2}\strain:{\bf C}:\strain\,dz \Big] dt$, where ${\bf C}$ denotes the fourth-order stiffness tensor. By employing the expressions for the strain components and expanding in $\kappa W \ll 1$ (and noting that $H \ll W$), it is straightforward to show that
\begin{equation}
\Pi = \Pi_s + \Pi_b + \Pi_g + {\mathcal{O}} (E H^3 W^3 \kappa^4, E H W^7 \kappa^6),
\label{eq1}
\end{equation}
where $\Pi_s = - (f_1 \kappa_1 + f_2 \kappa_2)WH/2$ denotes the contribution from the surface stresses, $\Pi_b = (\kappa_1^2 + \kappa_2^2 + 2\nu \kappa_1 \kappa_2) EWH^3/24(1-\nu^2)$ is the bending energy, and $\Pi_g = EHW^5 (\kappa_1 \kappa_2)^2/640(1-\nu^2)$ is the stretching energy arising from the geometric nonlinearity associated with non-zero Gauss curvature.  In general, terms associated with bending deformation are of the form $ \sim E W H^3 \kappa^2 (\kappa W)^{2n}$, where $n=0,1,2,3,...,$ while those associated with Gaussian curvature are of the form $ \sim E H W^5 \kappa^4 (\kappa W)^{2m}$, where $m=0,1,2,3,...$.    Also, we note that in the special case of spherical bending (i.e., $\kappa_1 = \kappa_2 = 1/\rho$), $\Pi_g = EHW^5/[640(1-\nu^2) \rho^4]$ to leading order, in agreement with classical F\"oppl-von K\'arm\'an theory \cite{Landau, Majidi_2008}.

Since $\Pi_b \sim EWH^3 \kappa^2 $ and $\Pi_g \sim EHW^5 \kappa^4 $, a simple scaling analysis suggests that when $W \ll \sqrt{H/\kappa}$, $\Pi_b \gg \Pi_g$, and the nonlinear geometric effect becomes negligible. That is, bistability is controlled by a dimensionless geometric parameter $\eta \equiv W \sqrt{\kappa/H}$ \cite{Armon_Science2011}, where the curvature $\kappa = \texttt{max}\{|\kappa_1|, |\kappa_2|\}$.  In the limit $\eta \ll 1$, $\Pi_g$ can be ignored, and the laminate develops into a saddle shape, as shown in Figs.~\ref{fig:Exp1} B and C (when $f_1 = - f_2$). In this case, applying the stationarity conditions ($\partial \Pi/\partial \kappa_i = 0$, $i$ = 1,2) to Eq.~(\ref{eq1}) and neglecting $\Pi_g$, we recover the analytical results  in \cite{Chen_APL2011}: $\kappa_1 = 6(f_1 - \nu f_2)/E H^2 \quad\mathrm{and}\quad \kappa_2 = 6(f_2 - \nu f_1)/E H^2$, with $\Pi^* = -3(f_1^2 - 2\nu f_1 f_2 + f_2^2)/2EH$. On the other hand, in the limit where $\eta \rightarrow \infty$, the system bifurcates into two stable, nearly cylindrical configurations, for which one of the principal curvatures becomes very small over the interior of the shell, as shown in Figs.~\ref{fig:Exp1} D and E and the supplementary movie.

The parameter $\eta$ is essentially the same as the one proposed by Armon et al.~\cite{Armon_Science2011}. While $\eta$ quantitatively captures the effect of the geometric nonlinearity, it contains an unknown parameter, the curvature -- significantly restricts its utility.  This deficiency is remedied as follows.  Since $\kappa \sim f/E H^2$ (where $f \equiv \texttt{max}\{|f_1|, |f_2|\}$), we define a new dimensionless parameter, $\zeta \equiv \sqrt{f/EH} W/H$, which involves only {\it a priori} known material and surface stress parameters. We demonstrate that this new parameter arises naturally by considering the case $f_1 = -f_2$ relevant for our experiments and those of Armon et al.~\cite{Armon_Science2011}, and derive an analytical expression for the bifurcation threshold.

Applying the stationarity conditions to Eq.~(\ref{eq1}) and setting $f_1 = -f_2$ yields $(\kappa_1 + \kappa_2)\left[\kappa_1 \kappa_2 + 80 (1 + \nu) H^2/3W^4\right] = 0$. Therefore, either $\kappa_1 + \kappa_2 = 0$ which corresponds to the saddle-like shape, or $\kappa_1 \kappa_2 = - 80 (1 + \nu) H^2/3W^4$, which, when real solutions exist, yields the bifurcated solutions. A bifurcation occurs when both $\kappa_1 + \kappa_2 = 0$ and $\kappa_1 \kappa_2 = - 80 (1 + \nu) H^2/3W^4$ are satisfied, implying that $\kappa_1 = -\kappa_2 = \kappa = -\sqrt{80(1+\nu)/3} H /W^2$ or $\eta = W \sqrt{\kappa/H} = \eta_c = [80 (1 + \nu)/3]^{1/4}$. Importantly, our prediction for the bifurcation threshold ($\eta_c = 2.51$ for $\nu=0.49$) is in excellent agreement with the experimental result, $\eta_c \approx 2.5$, found by Armon et al.~\cite{Armon_Science2011}. Moreover, stationarity conditions at the bifurcation imply that $\kappa_1 - \kappa_2 = 2 \kappa = 6f^* (1 - \nu^2)/EH^2$, or $\kappa = 3f^* (1 - \nu^2)/EH^2$. Thus, the bifurcation threshold can also be expressed as
\begin{equation}
\zeta_c \equiv \sqrt{\frac{f^*}{EH}} \frac{W}{H} =  \frac{ \eta_c}{\sqrt{3 \left(1-\nu^2 \right)}} =  \left[\frac{80}{27(1 - \nu) (1 - \nu^2)}\right]^{1/4}.
\label{eq:new}
\end{equation}
Equation (\ref{eq:new}), which is the central result of the analysis, quantitatively expresses the fundamental condition for bistability in terms of a dimensionless parameter which only depends on the driving force (surface stress), material properties, and ribbon dimensions. Although the bifurcation formally occurs at $\zeta=\zeta_c$, since the shape change only takes place gradually as $\zeta$ increases, we expect the bistable behavior to manifest only when $\zeta \gg \zeta_c$.

In comparing theoretical predictions with experiments, we find that in both Figs.~\ref{fig:Exp1}B and \ref{fig:Exp1}C, $\zeta$ is comparable to $\zeta_c \approx 1.5$ (in B  $\zeta = 1.5$; in C  $\zeta = 1.4$), and the sheets adopt saddle shapes, as expected.   On the other hand, increasing $\zeta$ to $3.1$ leads to bistable behavior characterized by cylindrical shapes, as shown in Figs.~\ref{fig:Exp1}D and \ref{fig:Exp1}E, respectively.  Moreover, for the cylindrical shapes, the measured radius of curvature ($13.1 \pm 0.5$mm) is in reasonable agreement with the theoretical prediction $EH^2/[6 f (1-\nu^2)] \approx 13.8$mm.

\begin{figure}[t]
\centerline{\includegraphics[height=2.75 in]{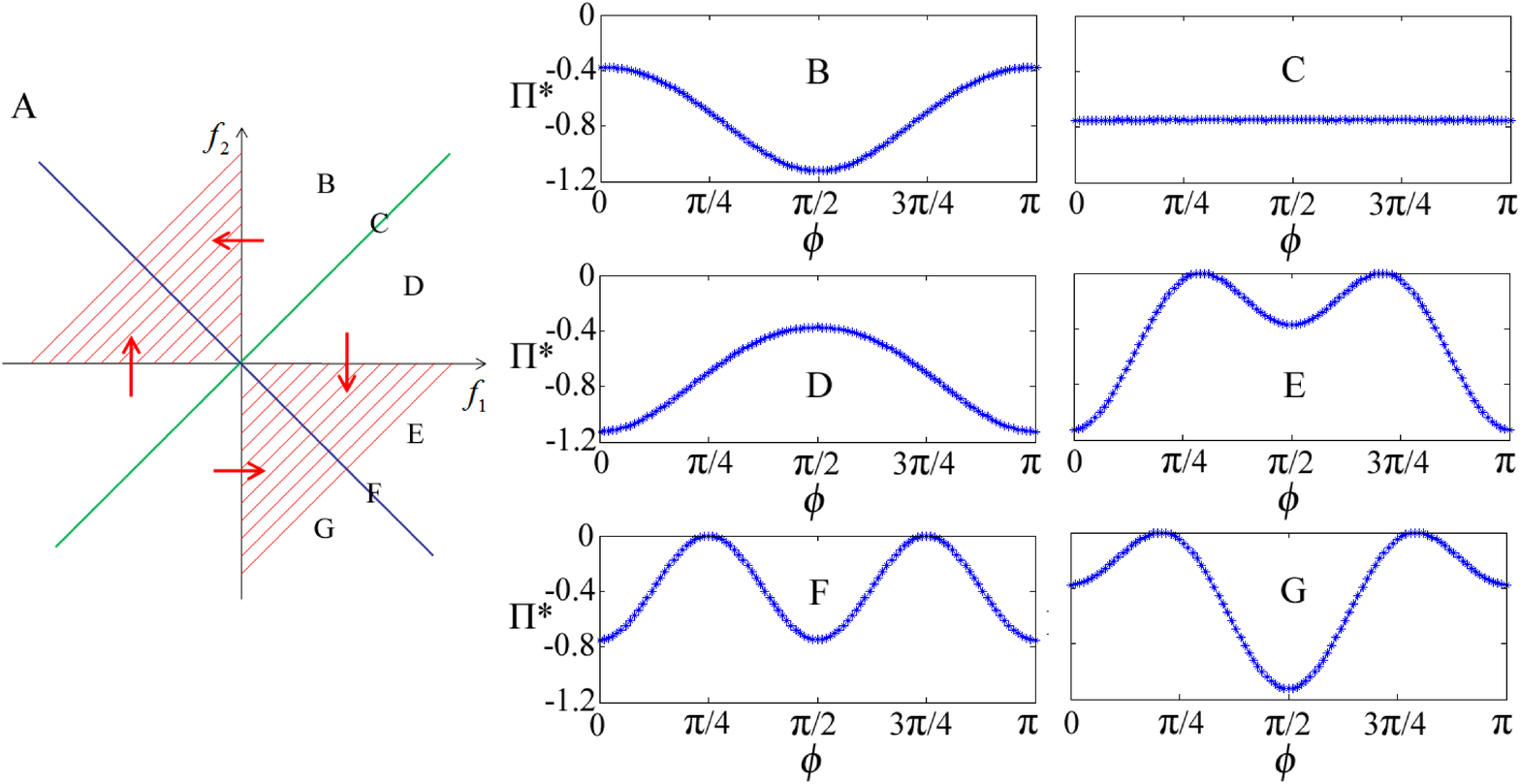}}
\caption{Bistability map in the $(f_1,f_2)$ space. In (A), the red shading indicates bi-stable regions while no shading indicates mono-stability. The red arrows denote transitions from mono-stable to bistable states. The total strain energy versus the orientation of the misfit axis for (B) $f_2/f_1 = \sqrt{3}$, (C) $f_2/f_1 = 1$, (D) $f_2/f_1 = 1/\sqrt{3}$, (E) $f_2/f_1 = -1/\sqrt{3}$, (F) $f_2/f_1 = -1$, and (G) $f_2/f_1 = -\sqrt{3}$.}
\label{fig:PiVSPhi}
\end{figure}

We now relax the condition $f_1 = -f_2$ and consider the role of surface stresses on mechanical instability in more general terms.  In the $\zeta \rightarrow \infty$ ($\eta \rightarrow \infty$) regime, the geometric nonlinearity requires that either $\kappa_1 \rightarrow 0$ or $\kappa_2 \rightarrow 0$ away from edges. Without loss of generality, we assume that $\kappa_y = \kappa_2 \rightarrow 0$. In order to examine stability, we first note that the principal bending axes ${\bf r}_1$ and ${\bf r}_2$ do not necessarily coincide with the geometric axes $\dx$ and $\dy$, but instead form an angle $\phi$ as shown in Fig.~\ref{fig:Exp1}A. In this case, the potential energy per unit length of the strip becomes $\Pi = - W H (f_1 C^2 + f_2 S^2)\kappa_1/2 + E H^3 W \kappa_1^2/24(1 - \nu^2)$, where $C \equiv \cos \phi$ and $S \equiv \sin \phi$.  Minimizing $\Pi$ with respect to both $\kappa_1$ and $\phi$ yields two sets of solutions:  (1) $\phi^* =0$ with $\kappa_1 = 6(1 - \nu^2)f_1/E H^2$ or (2) $\phi^* =\pi/2$ with $\kappa_1 = 6(1 - \nu^2)f_2/E H^2$.  Whether an extremum configuration ($\phi = 0$ or $\phi = \pi/2$) is locally stable depends on whether $\Pi''(\phi^*) \equiv d^2 \Pi/d \phi^2 |_{\phi=\phi^*} \geq 0$.  To this end, we write $\Pi''(0) = -6(1 - \nu^2)(f_2 - f_1)f_1 W/EH \propto -f_1^2(\beta-1)$ and $\Pi''(\pi/2) = -6(1 - \nu^2)(f_1 - f_2)f_2 W/EH \propto -\beta f_1^2(1-\beta)$, where $\beta \equiv f_2/f_1$ denotes another key dimensionless parameter.  Clearly, whenever $\beta < 0$, both $\Pi''(0)>0$ and $\Pi''(\pi/2)>0$,  implying the emergence of two mechanically stable configurations with one configuration in general being energetically more favorable than the other.  Interestingly, for $\beta=-1$, the two stable configurations are degenerate and have the same energy in the absence of edge effects. On the other hand, when $0<\beta<1$, $\Pi''(0)>0$ while $\Pi''(\pi/2)<0$, implying that the $\phi^*=0$ configuration is stable, while the $\phi^*=\pi/2$ one is mechanically unstable.  For $\beta>1$, the stabilities of the two extrema are reversed, with the $\phi^*=\pi/2$ configuration now displaying mechanical stability.  Finally, for the special case $\beta=1$, $\Pi^*(\phi)$ is constant, implying that the shell is in a neutrally-stable state \cite{Seffen_2011}.  In this case, the laminate can, in principle, curl into a nearly cylindrical shape with arbitrary orientation (at least in the absence of edge effects), and can also easily transition between these shapes drawn from a continuous family of neutrally-stable configurations.  The different possibilities are illustrated in Fig.~\ref{fig:PiVSPhi}.

Finally, when edge effects are taken into account, even when $\beta = \pm 1$, an asymmetry in dimensions ($L \neq W$) can result in the breaking of the degeneracy between the minimum energy configurations. As shown in Fig.~\ref{fig:Exp1} E, the cylindrical shape in the $\zeta \gg \zeta_c$ limit smoothly transitions into a more saddle-shaped one with finite Gauss curvature over a length scale $\sim W_c \sim \zeta_c H \sqrt{EH/f}$ as the ends of the cylinder are approached.  Within this transition region, the total potential energy per unit area can be approximated by $\Pi^*_1 = -3(f_1^2 - 2\nu f_1 f_2 + f_2^2)/2EH$, while for the same area in the cylindrical part, the energy per unit area is $\Pi^*_2 =  -3(1 - \nu^2) f_1^2/2EH$.  Since $\Pi^*_1 < \Pi^*_2$ for $\beta=\pm 1$, the edges reduce the overall deformation energy, and the most effective reduction is obtained for maximizing the overall edge length.  Thus, the laminate will curve along the long direction, in agreement with the recent 
results of Alben et al.~\cite{Alben_nanolett2011}.

In summary, large shell/plate deformation with geometric nonlinearity is treated through a novel, analytically tractable theoretical framework which combines continuum elasticity, differential geometry and stationarity principles. Two key dimensionless parameters ($\zeta \equiv \sqrt{f/EH} W/H$ and $\beta \equiv f_2/f_1$) were shown to govern structural bistability. It is noteworthy that $\zeta$ only involves the driving force (surface stress), material properties, and ribbon dimensions, in contrast to the geometric bifurcation parameter $\eta \equiv W \sqrt{\kappa/H}$ introduced by Armon et al.~\cite{Armon_Science2011}, which involves an {\it a priori} unknown parameter (curvature). On the one hand, whether a structure with linear elastic properties can exhibit bistability depends on its dimensions and curvatures ($\zeta \gg \zeta_c$, or $\eta \gg \eta_c$) due to nonlinear geometric effects. On the other, even when the necessary geometric condition is satisfied, the existence of bistability still depends on a mechanical parameter ($\beta < 0$). Our theoretical analysis also predicts the lifting of ground state degeneracy (when $\beta = \pm 1$) due to edge effects \cite{Alben_nanolett2011}. The non-linear geometric effects on bistability have been verified by table-top experiments, and our theoretical predictions for the bifurcation threshold are also in agreement with the experimental results of Armon et al.~\cite{Armon_Science2011}.

In a broader sense, our approach provides a means to quantify and understand the role of nonlinear geometric effects in multi-stable structures in a wide range of two-dimensional natural and engineered systems, such as morphogenesis and mechanical instability in biological systems, film-on-substrate electronics, and spontaneous coiling and buckling of strained thin films and ribbons.  Via reverse-engineering, it will also facilitate the design of multi-stable functional structures, from bio-inspired robots with smart actuation mechanisms to deployable, morphing structures in aerospace applications.

\begin{acknowledgments}
We thank Drs. Timothy Healey, Chi Li, and Clifford Brangwynne for helpful discussions, and Wanliang Shan for assistance with the experiments. This work has been supported, in part, by the Sigma Xi Grants-in-Aid of Research (GIAR) program,  National Science Foundation of China (Grant No.11102040), and American Academy of Mechanics Founder's Award from the Robert M. and Mary Haythornthwaite Foundation.
\end{acknowledgments}


\noindent
\baselineskip=15pt


\begin{thebibliography}{99}

\bibitem{Forterre_Nature2005} Y. Forterre, J. M. Skothelm, J. Dumais and L. Mahadevan, Nature {\bf 433}, 421 (2005).
\bibitem{Hyer_1981} M. W. Hyer, J. Compos. Mater. {\bf 15}, 296 (1981).
\bibitem{Daynes_2010} S. Daynes, C. G. Diaconu, K. D. Potter and P. M. Weaver, J. Compos. Mater. {\bf 44}, 1119 (2010).
\bibitem{Kebadze_2004} E. Kebadze, S. D. Guest and S. Pellegrino, Int. J. Solids Struct. {\bf 41}, 2801 (2004).
\bibitem{Vidoli_ProcRSocA2008} S. Vidoli and C. Maurini, Proc. R. Soc. A {\bf 464}, 2949 (2008).
\bibitem{Shahinpoor_2011} M. Shahinpoor, Bioinsp. Biomim. {\bf 6}, 046004 (2011).
\bibitem{Arrieta_APL2010} A. F. Arrieta, P. Hagedorn, A. Erturk and D. J. Inman, Appl. Phys. Lett. {\bf 97}, 104102 (2010).
\bibitem{Galletly_IntJSolStruct2004II} D. A. Galletly and S. D. Guest, Int. J. Solids Struct. {\bf 41}, 4503 (2004).
\bibitem{Guest_ProcRSocA2006} S. D. Guest and S. Pellegrino, Proc. R. Soc. A {\bf 462}, 839 (2006).
\bibitem{Seffen_ProcRSocA2007} K. A. Seffen, Proc. R. Soc. A {\bf 463}, 67 (2007).
\bibitem{Gigliotti_2004} M. Gigliotti, M. R. Wisnom and K. D. Potter, Composites Science and Technology {\bf 64}, 109 (2004).
\bibitem{Fernandes_2010} A. Fernandes, C. Maurini and S. Vidoli, Int. J. Solids Struct. {\bf 47}, 1449 (2010).
\bibitem{Armon_Science2011} S. Armon, E. Efrati, R. Kupferman, and E. Sharon, Science {\bf 333}, 1726 (2011).
\bibitem{Forterre_Science2011} Y. Forterre and J. Dumais, Science {\bf 333}, 1715 (2011).
\bibitem{Docarmo_DiffGeo} M. P. do Carmo, Differential Geometry of Curves and Surfaces, Prentice-Hall, Inc, Englewood Cliffs (1976).
\bibitem{Finot_JAP1997} M. Finot, I. A. Blech, S. Suresh and H. Fujimoto, J. Appl. Phys. {\bf 81}, 3457 (1997).
\bibitem{DomokosHealey_2004} G. Domokos and T. J. Healey, Int. J. Bifurcat. Chaos {\bf 15}, 871 (2005).
\bibitem{Biton_2007} Y. Y. Biton, B. D. Coleman and D. Swigon, J. Elasticity {\bf 87}, 187 (2007).
\bibitem{Kocsis_2012} A. Kocsis and D. Swigon, Int. J. Nonlinear Mech. {\bf 47}, 639 (2012).
\bibitem{Goriely_1998} A. Goriely and M. Tabor, Phys. Rev. Lett. {\bf 80}, 1564 (1998).
\bibitem{Huang_2012} J. Huang, J. Liu, B. Kroll, K. Bertoldi and D. Clarke, Soft Matter {\bf 8}, 6291 (2012).
\bibitem{Liang_PNAS2009} H. Y. Liang and L. Mahadevan, Proc. Nat. Acad. Sci. USA {\bf 106}, 22049 (2009).
\bibitem{Klein_Science2007} Y. Klein, E. Efrati and E. Sharon, Science {\bf 315}, 1116 (2007).
\bibitem{Dervaux_PRL2008} J. Dervaux and M. B. Amar, Phys. Rev. Lett. {\bf 101}, 068101 (2008).
\bibitem{Wyczalkowski2012} M. A. Wyczalkowski, Z. Chen, B. Filas, V. Varner and L. A. Taber, Birth Defects Res., Part C {\bf 96}, 132 (2012).
\bibitem{Suo_APL1999} Z. Suo, E. Y. Ma, H. Gleskova and S. Wagner, Appl. Phys. Lett. {\bf 74}, 1177 (1999).
\bibitem{Chen_JAM2004} X. Chen and J. W. Hutchinson, J. Appl. Mech. {\bf 71}, 597 (2004).
\bibitem{Rogers_PNAS2009} J. A. Rogers and Y. Huang, Proc. Nat. Acad. Sci. USA {\bf 106}, 10875 (2009).
\bibitem{Majidi_JMPS2010} C. Majidi, Z. Chen, D. J. Srolovitz and M. Haataja, J. Mech. Phys. Solids, {\bf 58}, 73 (2010).
\bibitem{Chun_nanolett2010} I. S. Chun, et al, Nano Lett. {\bf 10}, 3927 (2010).
\bibitem{Zhang_nanolett2006} L. Zhang, et al, Nano Lett. {\bf 6}, 1311 (2006).
\bibitem{Chen_APL2011} Z. Chen, C. Majidi, D. J. Srolovitz and M. Haataja, Appl. Phys. Lett. {\bf 98}, 011906 (2011).
\bibitem{Sawa_PNAS2011} Y. Sawa, et al, Proc. Nat. Acad. Sci. USA {\bf 108}, 6364 (2011).
\bibitem{Lee_JAP2011} H. Lee, M. Cho and B. Lee, J. Appl. Phys. {\bf 109}, 074314 (2011).
\bibitem{Shenoy_ACSNano2010} V. B. Shenoy, C. D. Reddy and Y. Zhang, ACS Nano {\bf 4}, 4840 (2010).

\bibitem{Footnote1} The latex sheets were produced by {\it Beijing Saili Physical Education and Development Inc.}, with thickness $0.25$mm and Young's modulus $1.65$MPa as measured by hanging weights on test strips.  The elastic strips were Acrylic, Wall-Mounting Tape, produced by {\it {Hongkong Golden Lion Inc.}} with thickness $1.0$mm and Young's modulus 8.5MPa. The Poisson's ratios of the acrylic strips and latex sheets were $0.37$ \cite{Powers_1972} and $0.49$ \cite{Kaazempur-Mofrad_2003}, respectively.

\bibitem{Powers_1972} J. M. Powers and R. M. Caddell, Polymer Engineering and Science {\bf{12}}, 432 (1972).

\bibitem{Kaazempur-Mofrad_2003} M. R. Kaazempur-Mofrad et al., Computers and Structures  {\bf{81}}, 715 (2003).

\bibitem{note} For a plane stress problem, the strain tensor $\gamma$ has the following components: $\gamma_{xx}(t,z) = \kappa_x z  + \kappa_1 R(t) - 1$, $\gamma_{yy}(t,z) = \kappa_y z$, and $\gamma_{zz}(t,z) = -\nu (\gamma_{xx} + \gamma_{yy})/(1-\nu)$,
where $t$ denotes the arclength along ${\bf r}_2$ measured from the big circle, $R(t)$ denotes the radius of the circle lying on the torus at a distance $t$ away from the big circle, and $\kappa_1 R(t) - 1$ is the extra tensile/compressive strain due to the geometric nonlinearity. For a torus, $\kappa_x$ and $\kappa_y$ denote the principal curvatures along ${\bf r}_1$  and ${\bf r}_2$ respectively: $\kappa_x = \cos (\kappa_2 t)/R(t)$, where $R(t) = 1/\kappa_1 + [\cos(\kappa_2 t) - 1]/\kappa_2$, and $\kappa_y = \kappa_2$, such that $\gamma_{xx}(t,z) = [\cos(\kappa_2 t)z]/ \left\{1/\kappa_1 + [\cos(\kappa_2 t) - 1]/\kappa_2 \right\}  + [\cos(\kappa_2 t) - 1]\kappa_1/\kappa_2$, $\gamma_{yy}(t,z) = \kappa_2 z $, and $\gamma_{zz}(t,z) = -\nu (\gamma_{xx} + \gamma_{yy})/(1-\nu)$.


\bibitem{Landau} L. D. Landau, E. M. Lifshitz, Theory of Elasticity, 3rd edn (Pergamon, 1986).
\bibitem{Majidi_2008} C. Majidi, R. S. Fearing, Proc. R. Soc. A {\bf 464}, 1309 (2008).
\bibitem{Seffen_2011} K. A. Seffen and S. D. Guest, J. Appl. Mech. {\bf 78}, 011002 (2011).
\bibitem{Alben_nanolett2011} S. Alben, B. Balakrisnan and E. Smela, Nano Lett. {\bf 11}, 2280 (2011).




\end{thebibliography}

\end{document}